\documentclass[11pt]{article}
\usepackage{epsfig}
\usepackage{color}

\begin{document}
\begin{titlepage}

\hfill {\today: hep-ph/}

\begin{center}
\ \\
{\Large \bf  The Inhomogeneous  Phase of Dense Skyrmion Matter}
\\
\vspace{.30cm}

Byung-Yoon Park$^{a}$, Won-Gi Paeng$^{b}$
\\ and Vicente Vento$^{c}$

\vskip 0.20cm

{(a) \it Department of Physics,
Chungnam National University, Daejon 305-764, Korea}\\
({\small E-mail: bypark@cnu.ac.kr})

{(b) \it AI Lab., Clunix, Seoul 07299, Korea}\\
({\small E-mail: wgpaeng@clunix.com})

{(c) \it Departament de F\'{\i}sica Te\`orica and Institut de F\'{\i}sica Corpuscular}\\
{\it Universitat de Val\`encia and Consejo Superior de Investigaciones Cient\'{\i}ficas}\\
{\it E-46100 Burjassot (Val\`encia), Spain} \\ ({\small E-mail:
Vicente.Vento@uv.es})

\end{center}
\vskip 0.3cm

\centerline{\bf Abstract}

It was predicted qualitatively  in ref.~\cite{PMRV02}  that skyrmion matter at low density is stable in an inhomogeneous phase 
where skyrmions condensate into lumps while the remaining space is mostly empty. The aim of this paper is to proof quantitatively
 this prediction.  In order to construct an inhomogeneous medium we distort the original FCC crystal to produce a phase of planar
  structures made of skyrmions. We implement mathematically these planar structures by means of the 't Hooft instanton solution
   using the Atiyah-Manton ansatz. The results of our calculation  of the average density and energy confirm the prediction suggesting 
    that the phase diagram of the dense skyrmion matter is a lot more complex than a simple phase transition from the skyrmion 
    FCC crystal lattice to the half-skyrmion CC one. Our results show that skyrmion matter shares common properties with standard 
    nuclear matter developing a skin  and leading to a binding energy equation which resembles the Weisz\"{a}cker mass formula.
\vskip 0.5cm

\vskip 0.3cm \leftline{Pacs: 12.39-x, 13.60.Hb, 14.65-q, 14.70Dj}
\leftline{Keywords: skyrmion, dense matter, phase transition}

\end{titlepage}
\section{Introduction}

The Skyrme model~\cite{Skyrme61} describes baryons as topological solitons 
of an effective meson Lagrangian, which can be interpreted 
as a large $N_c$ limit of QCD~\cite{Witten79}.
Once the topological winding number is identified with
the baryon number the model provides an excellent framework for hadron physics leading to a reasonable description of 
the properties of  single baryons~\cite{ANW83}, 
the baryon-baryon interaction~\cite{JJP85,VV85}, 
the baryon-meson interaction~\cite{SWHH89}, 
the properties of light nuclei~\cite{BS97, HMS98}.
and even of their excited states~\cite{MMW07}.

The baryon number one particles associated with this lagrangians are called skyrmions.
Due to its classical nature, which may be subject to quantization if required, 
the Skyrme model is also able to describe dense matter, the so called skyrmion matter,
free from any obstacles such as the ``sign problem" in lattice QCD~\cite{Klebanov85, GM87} . 
Moreover, the strong interaction generated by the skyrmions can be treated nonperturbatively. 
The great advantage of the model is that the same Skyrme Lagrangian 
can describe both infinite baryonic matter 
and its mesonic fluctuations  in a unified way~\cite{LPMRV03}.     

We have carried out several studies on dense matter with this scheme~\cite{PV10}. 
Our  interest has been in understanding the phase transitions of skyrmion matter 
and the associated changes in the hadron properties. 
As the density of the matter increases,
 skyrmion matter undergoes in some models a phase transtion 
from a FCC (face centered cubic) structure made of localized single skyrmions 
to a CC (cubic crystal) made of localized ``half-skyrmions" , i.e. that is skyrmion like particles with baryon 
number $1/2$ \footnote{
In other models the transition goes from single skyrmion CC to half-skyrmion BCC (body centered cubic).}.
In all cases, when matter is made of half-skyrmions, the value of the half-skyrmions at the lattice sites is such that 
the average value of  the non linear field $U = \exp(-i \vec{\tau}\cdot \vec{\pi}/f_\pi)$, vanishes.   
The vanishing of $\langle U \rangle$  was interpreted as a symptom 
of  chiral symmetry restoration in a dense medium. 
However, in ref.~\cite{LPMRV03}, it was pointed out that  $\langle U\rangle =0$ does not mean 
the vanishing of the pion decay constant in medium and therefore does not imply the restoration of chiral symmetry. 
The mechanism that takes place is well known under the name of ``pseudo-gap" scenario~\cite{HY01}.
At higher density $f_\pi$ does vanish and chiral symmetry is restored.
Thus the chiral restoration phase transition in the most naive Skyrme model happens in a two step process.
This unconventional realization is resolved by incorporating a dilaton field into the model Lagrangian, 
which is associated with the scale anomaly of QCD~\cite{LPMRV03}. 
Moreover, by incorporating more mesons such as  the $\rho$ and the $\omega$ 
into the model Lagrangian via a hidden local gauge symmetry~\cite{BKY88}, the description improves resembling closely the standard scenario~\cite{PRV04} leading to
 quantitative predictions on physical quantities such as the critical density of the phase transition~\cite{MHLOPR13}.  

However, a fundamental question on the phase transition remains.
From the very beginning the conventional approach adopted to study skyrmion matter 
assumes as ansatz for the minimum energy configuration a homogeneous crystal structure with certain symmetry~\cite{Klebanov85}. 
Although it is far from the Fermi liquid phase, which is generally accepted as a state for the normal nuclear matter, 
this solid crystaline structure is a reasonable starting point. 
However, the problem in this approach is that the phase transition occurs at a density 
where the pressure of the system is negative. 
Thus  the system is unstable against collapse and some mechanism should be provided to keep the system 
in that homogeneous phase. 
As suggested in ref.~\cite{PMRV02} at this density, where the system has negative pressure, the homogeneous 
symmetric configuration has no advantage in comparison to some  inhomogeneous phase where the skyrmions that are close to each other 
form lumps and leave large portions of space empty. This reasoning is supported by astrophysics where it is known that nuclear matter 
inside a neutron star appears in inhomogeneous phases 
known due to their structure by pasta names: gnocchi, spaghetti, lasagna, antispaghetti and
antignocchi~\cite{CH17}. 
In here, as a first step in the description of inhomogeneous skyrmion matter,
we construct mathematically and study a few  planar structures made of skyrmions. 

Contrary to the case of homogeneous skyrmion matter with adequate symmetries, 
the construction of  inhomogeneous skyrmion matter structures is not easy. 
In the case of a few skyrmions, one can  find the minimum energy configuration numerically~\cite{BS97},
or by means of useful mathematical tools such as the  rational map ansatz~\cite{HMS98}. 
In ref.~\cite{BS98}, a rational map ansatz is modified and applied to study a two-dimensional skyrmion lattice with hexagonal symmetry.  
In the present work, as in ref.~\cite{PMRV02}, we adopt the Atiyah-Manton ansatz~\cite{AM89} and construct a few layers of ``skyrmion sheets" as a model for inhomogeneous skyrmion matter.
We recall that in the Atiyah-Manton ansatz, a multi-skyrmion configuration can be generated from an instanton-like function in four dimensions. 
The instanton-like functions  provide an intuitive way of constructing complex skyrmion configurations 
by matching the singular points of the instanton-like function to the single skyrmion positions.  
By modifing the instanton-like function we can incorporate also the relative orientations to each skyrmion. 
By following this procedure we can construct a single sheet of skyrmions where one skyrmion is lodged in a particular lattice site 
and its nearest skyrmion neighbors  have relative orientation leading to the most attractive configuration.
We study in here single, double and triple arrays of sheets, which provide information 
on how the skyrmions behave in the bulk of a  matter lump and on its surface.

In the next section we describe our model lagrangian and the use of the Atiyah-Manton ansatz to construct skyrmion matter structures.
In Section 3 we describe how two construct two dimensional matter structures, which we call sheets, its symmetries and calculate average densities for the minimum average energy per baryon
for each of the structures. These sheets describe our model for an inhomogeneous phase. Finally in section 4 we draw some conclusions.

\section{Model Lagragian and Atiyah-Manton Ansatz}
Our starting point is the simplest  Skyrme Lagrangian~\cite{Skyrme61}, which reads
\begin{equation}
{\cal L} 
= \frac{f_\pi^2}{4} \mbox{Tr}\left( \partial_\mu U^\dagger \partial^\mu U \right)
+\frac{1}{32e^2} \mbox{Tr} \left[U^\dagger \partial_\mu U, U^\dagger \partial_\nu U\right]^2,   
\label{lag0}
\end{equation}
where $U =\exp(i\vec{\tau}\cdot\vec{\pi}/f_\pi) $ with the pion fields $\vec\pi$ and $SU(2)$ Pauli matrices $\vec\tau$. 
The model Lagrangian contains two parameters: the pion decay constant $f_\pi$ and the {\it Skyrme parameter}  $\;e$. 
By expressing the energy in units of $(6 \pi^2) f_\pi/e$ and the length in units of $1/e f_\pi$, 
the Lagrangian (\ref{lag0}) can be rewritten as 
\begin{equation} 
{\cal L} = - \frac{1}{24\pi^2}  \mbox{Tr}\left( L_\mu L^\mu \right)  
+ \frac{1}{192\pi^2} \mbox{Tr} \left[ L_\mu, L_\nu \right]^2 
\label{lag}
\end{equation}
where we have introduced the ``left-current" $L_\mu$ defined as 
\begin{equation}
L_\mu \equiv \partial_\mu U U^\dagger, 
\label{left_current}
\end{equation} 
$L_\mu$ are traceless and their spatial components  are of the form  $ L_i = \partial_i U U^\dagger,=i \vec\tau\cdot {\vec\ell_i}$,
which serves as definition of ${\vec\ell_i}$.  
The Faddeev-Bogomoln'y bound of the soliton solution carrying  baryon number (winding number) $B$ 
results simply as 
\begin{equation}
E \geq |B|, 
\label{bound}
\end{equation}
where $E$ is the energy of the static soliton 
\begin{equation}
E = \int d^3 x \left( {\cal E}_2 + {\cal E}_4 \right)
 =  \int d^3 x \left\{ \frac{1}{12\pi^2} {\vec\ell}_i \cdot {\vec\ell}_i
    + \frac{1}{24\pi^2} ({\vec\ell}_i \times {\vec\ell}_j )^2 \right\}, 
\label{E}
\end{equation}
and the baryon number is given by 
\begin{equation}
B = \int d^3 x \rho_B 
=  \varepsilon_{ijk} \int d^3 x \left\{ \frac{1}{24\pi^2} {\vec\ell}_i \cdot ({\vec\ell}_j\times {\vec\ell}_k) \right\}.
\label{B}
\end{equation}
For later convenience, we have introduced the densities ${\cal E}_{2,4}$ and $\rho_B$. 
The spherical hedgehog solution, i.e. the lowest energy solution for the $B=1$ skyrmion, solved exactly has an energy in our units of $E=1.23$, 
which is much greater than the bound. The lowest value found so far is $E/B=1.04$ in the half-skyrmion CC phase~\cite{PV10,PMRV02}.   

The Atiyah-Manton ansatz~\cite{AM89}, provides a skyrmion configuration carrying baryon number $B$ from the  
time component of the gauge potential of an instanton carrying the same charge in four dimensional Euclidean space.  
Explicitly, it reads
\begin{equation}
U(\vec x) = CS \left\{ {\cal P} \exp \left[ \int_{-\infty}^{\infty} -A_4 (\vec{x},t) dt\right] \right\} C^\dagger
\label{AM_ansatz}, 
\end{equation}
where $A_4 (\vec{x},t)$ is the time component of the gauge potential of an instanton of charge $N(=B)$. 
The integration is along the time direction  keeping the time-ordering as denoted by the symbol ${\cal P}$. 
In eq.(\ref{AM_ansatz}), $S$ is a constant $SU(2)$ matrix whose mission is to let  $U$ approach the vacuum value 1 
at infinity and $C$ is also an $SU(2)$  matrix that provides the global rotation of $U$ in the isospin space~\footnote{In the 
description of  skyrmion matter the value at infinity is taken at the points of the lattice chosen by the symmetry 
of the crystal so that the system has the proper baryon number.  For example, 
in the half-skyrmion phase, the points with $U=+1$ become the center of the half-skyrmions,  
while the original skyrmion centers with $U=-1$ remain as the center of the other half-skyrmion. }. 
By treating $C$ as collecive variables and quantizing its degrees of freedom, one may introduce  isospin to the system
and evaluate the symmetry energy of  matter~\cite{LPR11}.

An exact solution to $A_4 (\vec{x}, t)$ satisfying the equation of motion $\partial_\mu F^{\mu\nu} = 0$ 
together with the self-duality condition can be found in the form of 
\begin{equation}
A_4 (\vec{x},t) = \frac{i}{2} \vec\tau \cdot \vec\nabla (\ln\Phi), 
\label{A4_ansatz}
\end{equation}
with a scalar function $\Phi$. This ansatz reduces the equation of motion for $A_\mu$ to $\partial^2 \Phi = 0$~\cite{JNR77}. 
Various solutions to $\Phi$ carrying instanton charge $N$ are available and each yields different soliton 
configurations. The one that better suits our purposes is  't Hooft's instanton solution~\cite{tHooft,CF77}, 
where $\Phi$ is given by
\begin{equation}
\Phi = 1 + \sum_{n=1}^B \frac{\lambda_n^2}{(x-X_n )^2}.
\label{Phi}
\end{equation}
When this $\Phi$ is substituted into eq.(\ref{AM_ansatz}) with enough separations between the instantons, 
the spatial components of $X_n$, namely the singular points of the instanton solution, become  the positions 
of the centers of the skyrmions. Thus, this solution provides an intuitive way of arranging the skyrmions in  
three dimensional space.   

Two neighboring skyrmions are in the most attractive configuration 
when they have a specific relative orientation: one skyrmion is rotated in isospin space with respect to the other an angle $\pi$ 
with respect to an axis that is perpendicular to the line joining their centers.
In order to incorporate such adequate orientations for each the skyrmions we use a modified $A_4(\vec{x},t)$ 
developed in the liquid instanton model~\cite{CDG78}. There  Eqs.(\ref{A4_ansatz}) and (\ref{Phi}) are modified to
\begin{equation}
A_4(\vec{x},t) = \frac{i}{2} \sum_{n=1}^B C_n \vec{\tau} \cdot\vec{\nabla}
 (\ln\phi(x,X_n, \lambda_n)) C_n^\dagger,
\label{A4_mdf}
\end{equation}
and 
\begin{equation}
\phi(x, X_n, \lambda_n ) = 1 + \frac{\lambda_n^2}{(x-X_n)^2}.
\label{phi_small}
\end{equation}
The change has been motivated because the rotation does not act on the scalar function but does act on $\vec\nabla \phi$. 
Although eq.(\ref{A4_mdf}) is no more an exact solution of the equation of motion for the gauge fields,
the homotopy still provides us a $U$ with the correct baryon number.  
What plays the most important role in this solution are the $N$ singularities in eq.(\ref{AM_ansatz}) since they characterize the lattice structure mathematically. 
The $SU(2)$ rotation for each instanton done by $C_n$ correspond exactly to the rotations for each skyrmion in the skyrmion configuration.

\begin{figure}
\begin{center}
  \includegraphics[width=\linewidth]{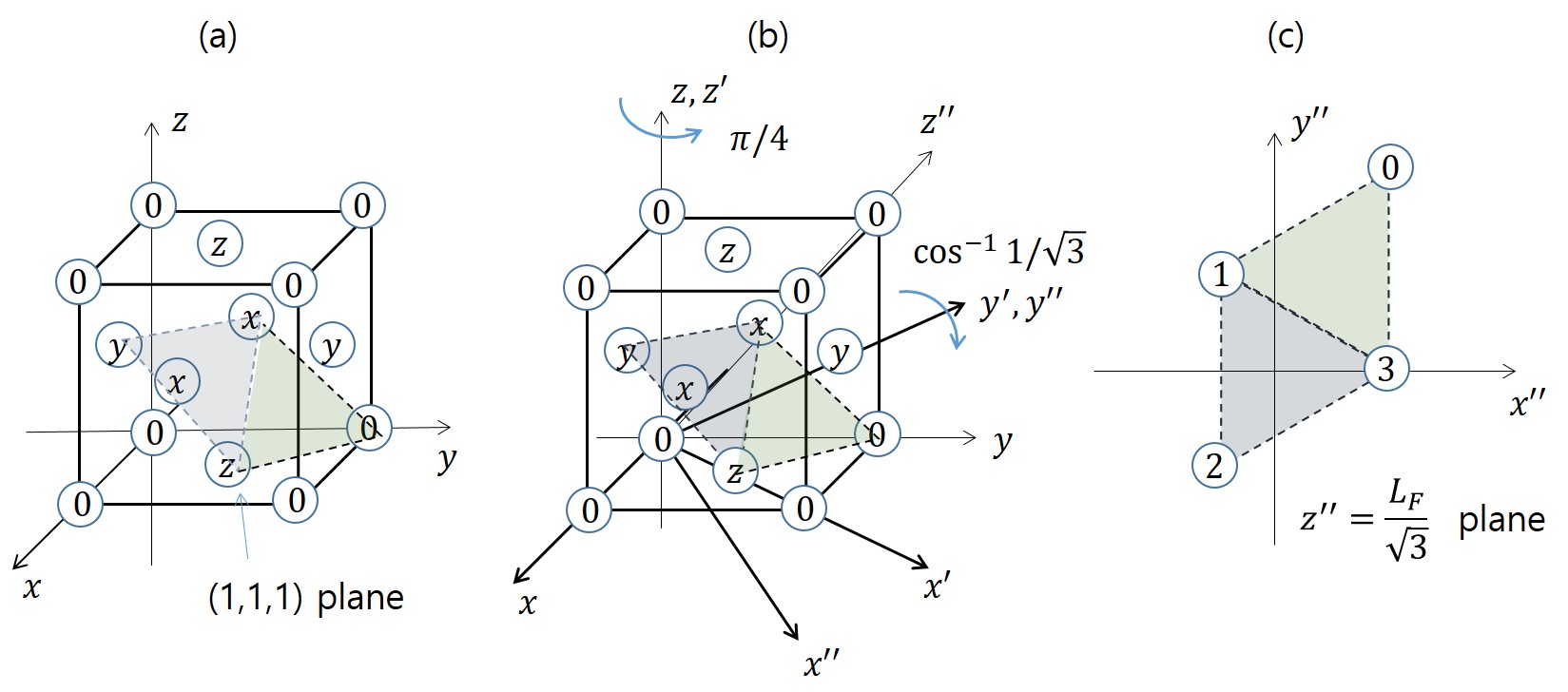}
  \caption{(a) The most attractive arrangement of skyrmions in an FCC cell. The unrotated skyrmions are occupying each of the vertices 
of the cubic cell. The skyrmions denoted by $x$, $y$, $z$ are located at the center of the sides parallel to the  $y-z$, $z-x$, $x-y$ planes 
repectively,  and they are  rotated by an angle $\pi$ with respect to the axis of the label. 
(b) Euler rotation of the coordinates to let the (1,1,1) plane be parallel to the new $(x^{\prime\prime},y^{\prime\prime})$ plane. 
(c) Arrangement of the skyrmions in a unit cell of the triangular lattice. 1, 2, 3 denote the skyrmions rotated by an angle $\pi$ 
with respect to the axis $x$, $y$, $z$  in the {\em  old} coordinate system respectively.  }
\end{center}
\end{figure}
In ref.\cite{PMRV02}, the Atiyah-Manton ansatz with the modified instanton configuration was applied to construct an
homogeneous FCC crystal. As shown in Fig.~1, to to have the most attractive relative orientations we have to locate skyrmions at each 
site of the FCC crystal in a specific way: the unrotated skyrmions (denoted by 0) are put at the vertices of a  cubic cell and the ones 
rotated by an angle $\pi$ with respect to an axis (denoted by $x, y, z$) 
perpendicular to the correponding sides of the cube are put at the centers of the corresponding sides.
The advantage of the ansatz  $A_4(\vec{x},t)$ defined as in eqs.(\ref{A4_mdf}) and (\ref{phi_small}) is that this structure can be 
mathematically implemented by taking the space components of $X_n$ at skyrmion sites  and their orientations by $C_n$. 
 When the size of the cell is large enough, that is, at low densities, the resulting skyrmion configuration 
has just the FCC structure made of individual single skyrmions arranged in cells as in Fig.~1. As we increase the density beyond the  critical value, 
 matter has a lower energy per baryon in a different configuration. The crystal structure becomes CC as each skyrmion  of the previous configuration 
 is separated into two well localized lumps which are almost half-skyrmions  (B=1/2).  The fact that they are not exactly half-skyrmions has to do with 
 the limitations of using an approximate ansatz.

As already mentioned the problem with  this phase transition is that it occurs at a densiy where matter has negative pressure.
The system, instead of remaining  in an unstable homogeneous state, prefers to form stable lumps of higher density  with 
large portions of space empty, i.e. a distribution of matter resembling a thin soup of pasta.
It is known that in the core of neutron stars, depending on the density of nuclear matter, several 
inhomogeneous phases of this type have been suggested \cite{CH17}. 

Let us model some inhomogeneous phases  in skyrmion matter to guide ulterior massive numerical computations.
We may get some insight in the inhomogeneous phases by distorting the cell dimensions in Fig.~1. 
For example, by enlarging the cell dimension in the $x$ and $y$ directions 
while keeping that in the $z$ direction, we construct strings of skyrmions, while if we enlarge the cell dimension only in $z$ direction, 
we will obtain skyrmion sheets, where skyrmions are lodged in square lattice sites and one skyrmion has four nearest neighbors. 
The more energetically favorable planar structures are obtained 
by taking the (1,1,1) plane shown explicitly in Fig.~1. There the skyrmions are in triangular lattice sites 
and each of them has six nearest neighbors with the most attractive relative orientations. 
The transition from the BCC to FCC associated with the (1,1,1) plane has been studied in ref.\cite{CJJVJ89} 
to describe the three dimensional structural phase transition. However, our main interest here is  the the study of the two dimensional structure itself.

With our Atiyah-Manton based approach using eqs.(\ref{AM_ansatz}) and (\ref{A4_mdf}), 
we can obtain such linear or planar structures rather simply by taking the space components of $X_n$ 
as the lattice sites for lines or  planes and by choosing proper $C_n$  to characterize the rotations.   
The resulting dimensional hexagonal skyrmion lattice has been studied in ref.\cite{BS98} using rational Donaldson maps by solving the full equations 
leading to the lowest energy configuration.  We will take their results as a reference value for our study 
and will go beyond to describe double and triple layers of planes which we call skyrmion sheets.

\section{Models for a inhomogeneous phases in Skyrme matter}

We are going to construct different inhomogeneous phases based on parallel skyrmion sheets leading to, given our previous culinary methaphor, a Lasagna type phase.
For convenience, we rotate the coordinate system so that the (1,1,1) plane in Fig.~1(a) 
is taken as $(x,y)$ plane in the new coordinate system. As shown explicitly in Fig.~1(b)
this can be done by a two step Euler rotation. 
At first,  we rotate by an angle $\pi/4$ about the $z$ axis
which yields new coordinates system $(x^\prime, y^\prime, z^\prime=z)$. 
Next,  we  rotate by an angle $\cos^{-1}(1/\sqrt{3})$ about the $y^\prime$ axis to get the coordinates system
$(x^{\prime\prime}, ~y^{\prime\prime}=y^\prime, ~ z^{\prime\prime})$.  
Explicitly, the corresponding rotation matrix is 
\begin{equation}
R = 
\left( 
\begin{array}{ccc} 
-\frac{1}{\sqrt3} & 0 & +\frac{\sqrt2}{\sqrt3} \\
0 & +1 & 0 \\
+ \frac{1}{\sqrt3} & 0 & +\frac{1}{\sqrt3} 
\end{array}
\right)
\left(
\begin{array}{ccc} 
+\frac{1}{\sqrt2} &  +\frac{1}{\sqrt2} & 0  \\
+\frac{1}{\sqrt2} & -\frac{1}{\sqrt3} & 0 \\
0 & 0 & 1 
\end{array}
\right)
= 
\left(
\begin{array}{ccc} 
+\frac{1}{\sqrt6} &  +\frac{1}{\sqrt6} & -\frac{\sqrt2}{\sqrt3}  \\
-\frac{1}{\sqrt2} & +\frac{1}{\sqrt2} & 0 \\
+\frac{1}{\sqrt3} & +\frac{1}{\sqrt3} & +\frac{1}{\sqrt3} 
\end{array}
\right)
\end{equation}
Fig.~1(c) shows how the skyrmions are arranged in the new $z^{\prime\prime}=L_F/\sqrt{3}$ plane. 
The $x,~y,~z$ axes in the old coordinate system are now transformed to new axis parallel to the unit vectors, 
whose components are 
\begin{equation}
\hat{n}_1 = \left( \begin{array}{c} +\frac{1}{\sqrt6} \\ -\frac{1}{\sqrt2} \\ +\frac{1}{\sqrt3} \end{array} \right), ~
\hat{n}_2 = \left( \begin{array}{c} +\frac{1}{\sqrt6} \\ +\frac{1}{\sqrt2} \\ +\frac{1}{\sqrt3} \end{array} \right), ~
\hat{n}_3 = \left( \begin{array}{c} -\frac{\sqrt2}{\sqrt3} \\0 \\ +\frac{1}{\sqrt3} \end{array} \right), 
\end{equation}
respectively. The rotations by an angle about these new axes in isospin space can be obtained by 
$C_i = i \vec\tau\cdot \hat{n}_i (i=1,2,3)$ and we  denote the rotated skyrmions by 1, 2 and 3. 
From now on, we perform our calculations always in the new coordinate system and drop the primes.

\subsection{Inhomogeneous phase of a single skyrmion sheet}
As discussed above, the minimum energy configuration 
of the skyrmions in a sheet-like structure is the one of the (1,1,1) plane.
Fig.~2(a) explicitly shows the location of skyrmions, from which we read the 
translational symmetries as
\begin{equation}
 U(x,y+L,z) = (\vec\tau\cdot \hat{n}_3) U(x,y,z) (\vec\tau\cdot \hat{n}_3), 
 U(x+\frac{\sqrt3}{2}L, y+\frac12 L,z) = (\vec\tau\cdot\hat{n}_1) U(x,y,z) (\vec\tau\cdot\hat{n}_1).
\label{trans_sym}
\end{equation}

It is straightforward to formulate this skyrmion arrangement in our approach:
we take the spatial components of $X_n$ in eq.\ref{A4_mdf}) 
as the lattice points and the rotations in the isospin space $C_n$ as shown in Fig.2(a).   

\begin{figure}
\begin{center}
  \includegraphics[width=\linewidth]{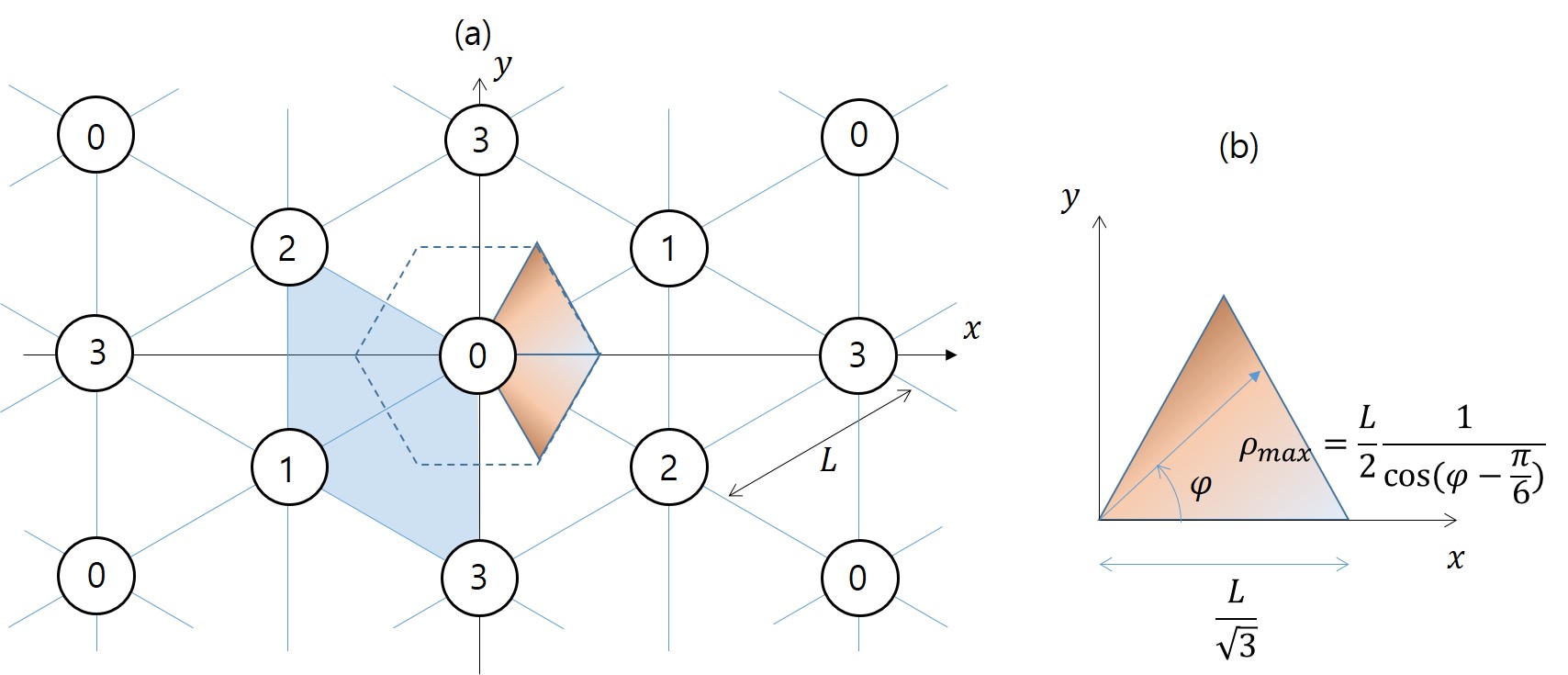}
  \caption{(a) Lattice points for the spatial components of $X_n$ and the relative orientations in the most attractive planar structure of skyrmions. 
A unit cell  is denoted by a shaded diamond and the hexagonal prism (unbound in $z$ direction) is shown as a hexagon. 
The 2-fold symmetry about the x-axis is shown explicitly. 
We denote the distance between the centers of two nearest neighboring skyrmions as $L$.
In terms of the FCC cell dimension $L_F$, $L=L_F/\sqrt2$.
(b) We show $\rho_{max}$ appearing in eqs.(\ref{B_box}) and (\ref{E_box}).  }
\end{center}
\end{figure}

Note that the summation in eq.(\ref{A4_mdf}) runs 
over all the triangular lattice sites $\vec{X}_n$ that expand to infinity. 
However, the summation converges since for $X_n \gg L$ 
\begin{equation}
\partial_i \ln\phi \sim \frac{(x-X_n)_i}{(x-X_n)^4}.
\label{asmpt}\end{equation}
and 
\begin{equation}
\sum_n \partial_i \ln \phi \sim \int d^2 X \frac{1}{|x-X|^3}. 
\end{equation}
Furthermore, the symmetry arising from the lattice structure provides one order higher suppression in the 
large $X_n$ contributions to the summation since, as can be seen in 
Fig. 2(a), the skyrmions 0, 1, 2, 3 are located radially opposite with respect to the skrymion 0 
at the origin. These symmetry make the leading order term in eq.~(\ref{asmpt}) $X_{n,i}/(x-X_n)^4$ vanish~\footnote{This symmetry makes the 
summation converge even in three dimensional lattice.}.

There are two more parameter sets in our ansatz:  $T_n$, the time component  of  $X_n$ and the instanton size parameters $\lambda_n$. 
To satisfy the symmetries (\ref{trans_sym}), $T_n$ and $\lambda_n$ should not depend on $n$. 
Let's denote these two numbers  $T$ and $\lambda$. 
$T$ can be absorbed in the integration process over $t$ in Eq.(\ref{AM_ansatz}) 
by changing to $t^\prime= t-T$ as the new variable.

In this single sheet configuration, the matter containing a single baryon  is located inside a hexagonal prism 
(unbound in $z$ direction) denoted by the dashed line. 
This arrangement yields the densities in the integrals of eqs.~(\ref{E}) and (\ref{B})
with 3-fold symmetry with respect to $2\pi/3$ rotations about the $z$ axis 
and a 2-fold symmetry with respect to a rotation by $\pi$ about  the $x$ axis. 
There is no reflection symmetry about the $z=0$ plane.
Thus, we need to integrate only over the triangular prism of $0\leq z < \infty$. 
In  cylindrical coordinates $(\rho, \varphi, z)$, the baryon number and the energy per baryon  
can be expressed  as  
\begin{equation}
1 = 12 \int_0^\infty \hskip-1em  dz \int_{\triangle} dxdy \rho_B 
=  12\int_{0}^\infty\hskip-1em dz \int_0^{\pi/2} \hskip-1em d\varphi \int_0^{\rho_{max}} \hskip-1em \rho d\rho \rho_B,
\label{B_box}
\end{equation}  
and 
\begin{equation}
E/B = 12 \int_0^\infty\hskip-1em dz \int_{\triangle} dxdy ( {\cal E}_2 + {\cal E}_4) 
=  12\int_{0}^\infty \hskip-1emdz \int_0^{\pi/2} \hskip-1em d\varphi \int_0^{\rho_{max}} \hskip-1em \rho d\rho 
({\cal E}_2 + {\cal E}_4 ),
\label{E_box}
\end{equation}  
with $\rho_{max}$ given by
\begin{equation}
\rho_{max} = \frac{L}{2} \frac{1}{\cos(\varphi - \frac{\pi}{6})}.
\end{equation}

Our problem becomes to minimize this eq.~(\ref{E_box}) by adjusting the distance $L$ 
and the size parameter $\lambda$. 
However, we can reduce our problem to a single parameter minimization. 
Let's rescale all the lengths by $L$
\begin{equation}
x_i = L \tilde{x}_i, ~~t = L t ~~ X_{n,\mu} = L \tilde{X}_{n,\mu}, ~~ \lambda = L \tilde{\lambda}.
\end{equation}
The function $\phi$ in (\ref{A4_mdf}) remains exactly of the same form under this scaling, %
\begin{equation}
\phi = 1 + \frac{\tilde{\lambda}_n^2}{(\tilde{x}-\tilde{X}_n)^2}.
\end{equation}
Thus the  $L$ dependence of the integrals in eq.(\ref{E_box}) becomes explicit as
\begin{equation}
E/B = L \left ( \tilde{E}/B \right)_2 + \frac{1}{L} \left( \tilde{E}/B \right)_4,
\label{EperB_scale}
\end{equation}  
where $ (\tilde{E}/B)_{2,4}$ are the integrals of the densities ${\cal E}_{2,4}$ evaluated with the tilded variables. 
We can now do the minimization with respect to the two parameters as follows.
We  evaluate $ (\tilde{E}/B)_{2,4})$ for a given $\tilde{\lambda}$, thereafter we minimize easily
eq.(\ref{EperB_scale}) by varying $L$, which leads us to  
\begin{equation}
(E/B)_{min} = 2\sqrt{(\tilde{E}/B)_{2}(\tilde{E}/B)_{4}}, ~~
L = \sqrt{(\tilde{E}/B)_{4}/(\tilde{E}/B)_{2}}
\end{equation}
with the value of $\lambda$ as $L\tilde\lambda$. 
Thus, all the expressions are now written in terms  of  a unique  parameter $\tilde{\lambda}$ with respect to which we have to vary.

\begin{figure}
\begin{center}
  \includegraphics[width=0.7\linewidth]{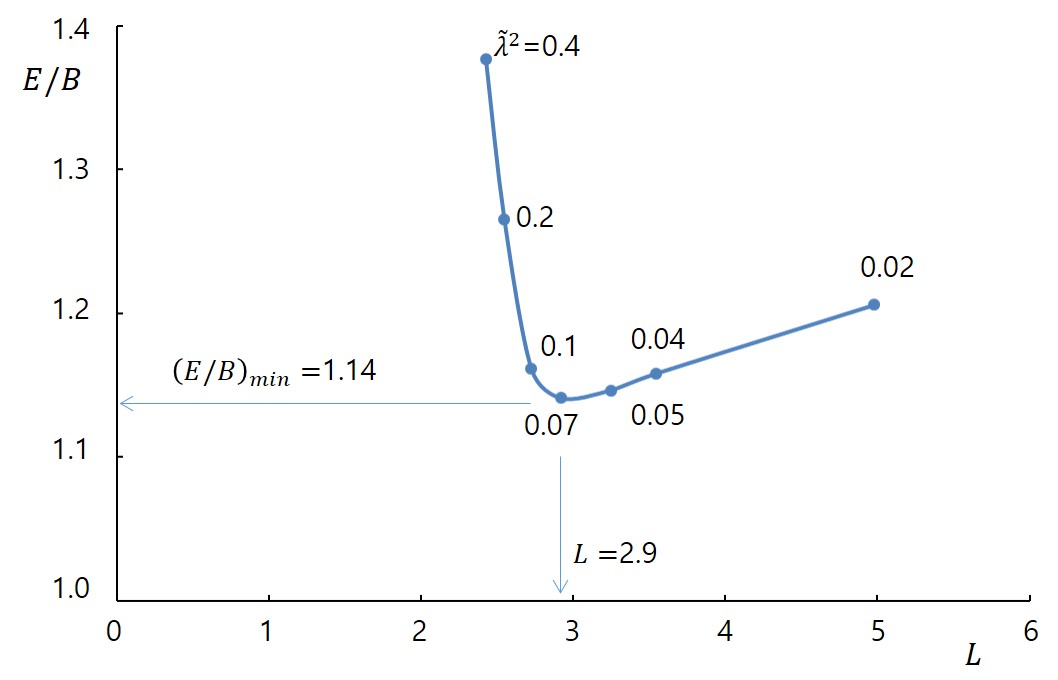}
  \caption{$(E/B)$ of a single skyrmion sheet as a function of $L$. The ${\tilde\lambda}^2$ values for obtaining  each points are shown too. 
The minimum value of $(E/B)$ is 1.14 at $L=2.95$.   }
\end{center}
\end{figure}

We show in Fig.~3  the results of such a minimization process. 
The $\tilde{\lambda}$ values used in the calculation are given at each point. 
The minimum energy per baryon is $E/B = 1.14$ at $L= 2.95$. 
Although this $(E/B)_{min}$ is larger than 1.076 of the exact solution to hexagonal lattice~\cite{BS98}, 
it is not too bad if we take into account that it is obtained by varying just  a single parameter.   
We may obtain better values by modifying the instanton-like function $\phi$ in eq.(\ref{phi_small}).    

It is reasonable to compare our result with those of ref.\cite{PMRV02} obtained by using the same Atiyah-Manton ansatz.
There, the phase transition from the single skyrmion FCC to half-skyrmion CC occurs when  
the energy per baryon  is  $(E/B) \sim 1.16$. 
That is, our two dimensional planar structure has lower $(E/B)$ than 
any homogeneous FCC and even the half-skyrmion CC near the critical density. 
We infer that skyrmion matter prefers to be in an inhomogeneous phase, 
where skyrmions condense to form planar structures surrounded by empty space.

\begin{figure}
\begin{center}
  \includegraphics[width=\linewidth]{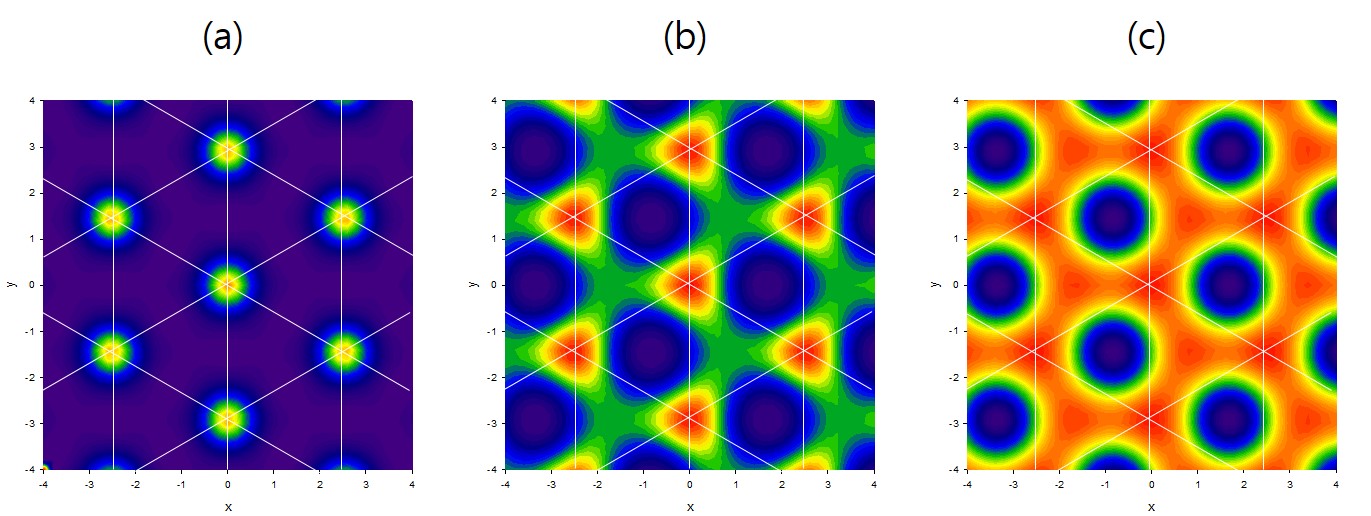}
  \caption{Baryon number distributions of the minimum energy configuration on the plane (a) $z=0$, (b) $z=0.5$ and (c)$ z=1.0$.
In each graph, the color of the contours denote relative densities only. The white lines show the triangular lattice.}
\end{center}
\end{figure}

In Fig.4, the minimum energy configuration is depicted by contour plots of baryon number 
densities on the plane (a) $z=0$, (b) $z=0.3$ and (c) $z=1.0$.
In each plot, the color of contours denote only  relative densities. 
The triangular lattice is explicitly shown by the white lines. 
At the central $z=0$ plane, matter is concentrated in narrow regions around the lattice points in the form of circular contours.
As $z$ increases, the density distribution expands to a wider region and the contour shape is distorted to curved triangles. 
Finally, in the tail region far from the central plane, the contours of higher density 
become hexagons, similar to the results of ref.~\cite{BS98}. 

Three vertices of the hexagon are the lattice points of the triangular lattice, while the other three 
are the centers of the triangles.  Matter gets accumulated in these full regions by the overlapping of the tails of the skyrmions. 
Although this overlapping happens in the low density tail region, it shows an interesting feature of the 
skyrmion approach, namely that it can describe naturally shape distortion of objects due to the interaction of neighbors

\begin{figure}
\begin{center}
  \includegraphics[width=0.8\linewidth]{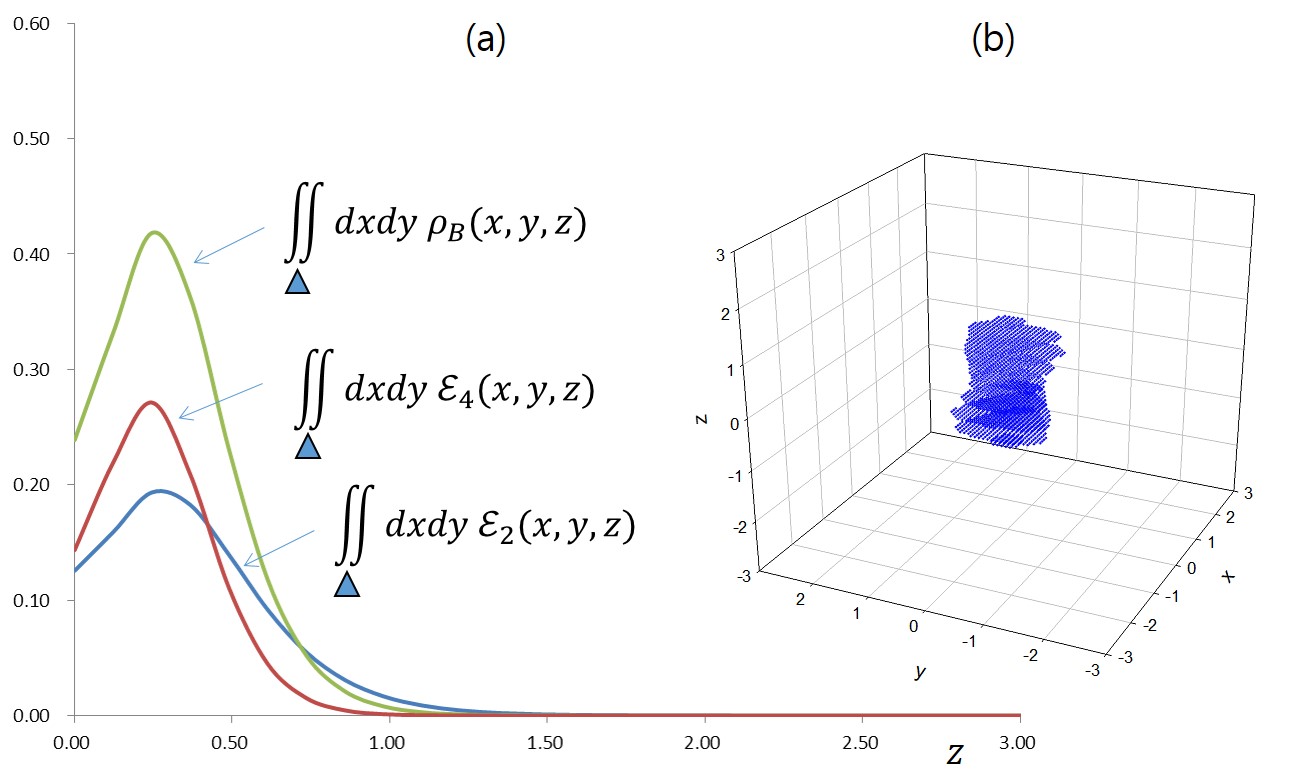}
  \caption{(a) Averaged density distributions along the $z$ direction. 
(b) A rough three dimensional plot of baryon number distribution in a hexagonal prism.  
We draw the positions where the density is larger than 10\% of the maximum values. }
\end{center}
\end{figure}

In order to see the density distributions along $z$ direction, in Fig.~5(a), 
we present  the densities integrated over the triangle region for a given $z$. 
They can be interpreted as the averaged densities over that area. 
Contrary from  what we had expected, they do not have maximum values at $z=0$. 
These results tell us that skyrmions in our minimum energy configuration are not  single objects sitting 
at each triangular lattice points. As can be seen in the rough three dimensional plot in Fig.~5(b), 
a single skyrmion is distributed into what looks as two recognizable objects. 
These two distinguishable density distributions may be taken as primogenial half-skyrmions.

\subsection{Inhomogeneous phase of a double layer of skyrmion sheets}
Now, let us consider two parallel skyrmion sheets. 
It is reasonable to assume that the minimum energy configuration is made by two nearest (1,1,1) planes of an FCC crystal.
It can be constructed by arranging skyrmions as shown in Fig.~6. 
In order to make nearest neighboring skyrmions at two layers equally separated, for example,  
0 in the lower layer and 1, 2, 3 in the upper layer (shown in the figure by shaded circles), 
the distance between the two central planes is $d=L/\sqrt{6}$. 
If we place the $z=0$ plane at the center between the two sheets,  
the matter carrying baryon number one  is located
inside the semi-infinite hexagonal prism defined by  $z\geq 0$ or $z\leq 0$. 

\begin{figure}
\begin{center}
  \includegraphics[width=0.7\linewidth]{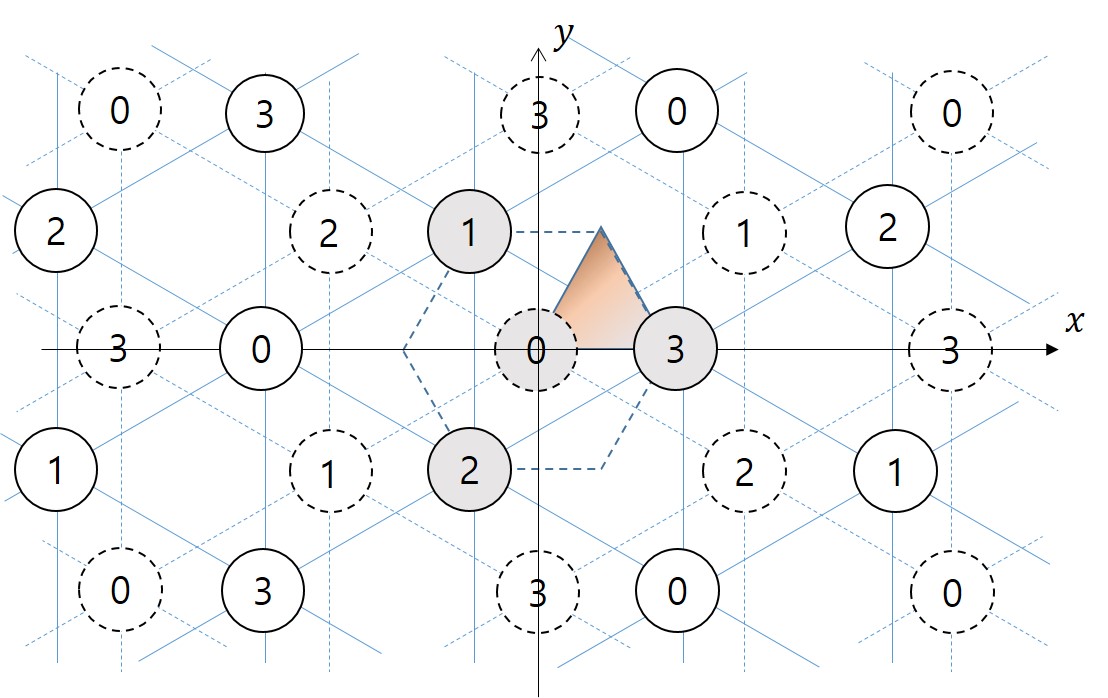}
  \caption{Arrangement of skyrmions in double layers. The skyrmions on the lower (upper) layer are denoted by dashed (solid) circles.}
\end{center}
\end{figure}

A similar numerical procedure as described in detail before for the single sheet phase leads to $(E/B)_{min}=1.11$ at $L=3.4$. 
The interaction with three more nearest neighbors in the other sheet reduces $E/B$ to 1.11 from 1.14 of the single skyrmion sheet.   
Compared to the reduction from 1.24 from the spherical hedgehog solution using the Atiyah-Manton ansatz, to 1.14 going to the 
single sheet with six nearest neighbors, this reduction is rather small.
Fig.~7 provides information on the minimum energy configuration, where we show
(a) the averaged densities over the triangular region for a given $z$ (not scaled) and (b) a  three dimensional plot where  
the region with  density greater than 10 \%  the maximum value. 
The dashed line in Fig.~7(a) denotes one of the lattice planes, $z/d=\pm 0.5$. 
The averaged densities outside the lattice plane show great similarity to those for the single sheet in Fig.~5(a). 
It may look natural, because we just stack the instanton-like objects on two layers. 
However, if we take into account the fact that the instanton-like function $\phi$ in eq.(\ref{A4_mdf})
is  long-range, it seems quite unnatural that the other layer does not affect the density distributions 
in this region. 
In Fig.~7(a), the baryon number in the hexagonal prism defined by  $0\leq z\leq d/2$ is exactly 0.5 and  
that in the semi-infinite hexagonal prism defined by $z\geq d/2$ is the remaining half baryon number. 
On the other hand, the energies in each hexagonal prisms are 0.53 and 0.58, respectively. 
This implies that the ``half"-skyrmions in between two lattice planes carry less energy than those 
outside the lattice planes. It is also interesting to see that two times 0.53, that is, 1.06  is  the
$(E/B)$ of the half skyrmion CC obtained using the Atiyah-Manton ansatz in ref.~\cite{PMRV02}. 
Furthermore, twice of 0.58 is close to what we have obtained as $(E/B)_{min}$ of single skyrmion sheet. 

In Fig.~7(b), besides the lumps around the lattices points, one can see newly generated lumps at the shaded $z=0$ plane. 
Although they are not fully symmetric\footnote{In ref.\cite{PMRV02}, 
the Atiyah-Manton ansatz could not produce perfectly symmetric half-skyrmions 
even in the FCC configuration.}, these generated lumps can be interprted as half-skyrmions.
That is, we have arranged the instanton-like objects on two sheets, but we end up with a triple layer of  ``half"-skyrmions. 
Of course, in the configuration obtained with larger $L$ (and larger $(E/B)$ consequently), the lumps appear only around the 
lattice points of the two sheets.

Our double layer results provide enough informations on how the skyrmions~\footnote{We should say half-skyrmions 
instead of skyrmions, because there is no inside skyrmion in the double layer configuration. Only once  they separate into 
half-skyrmions we can distinguish between inside the layer objects and those on the surface.}  
behave inside the dense matter and on its surface. 
That is, the lumps inside matter have all the possible nearest neighbors and have lower energy, while those on the surface 
have less neighbors and larger energy. This discussion resembles the liquid drop model arguments, i.e. common statements to all matter made 
of particles interacting with strong but short-range forces. In our case these interactions are just a consequence of the 
Lagrangian eq.(\ref{lag0}) even with the massless pions. Furthermore, it is extremely interesting that we have obtained 
such a result by using the long-range instanton-like function $\phi$.

\begin{figure}
\begin{center}
  \includegraphics[width=0.8\linewidth]{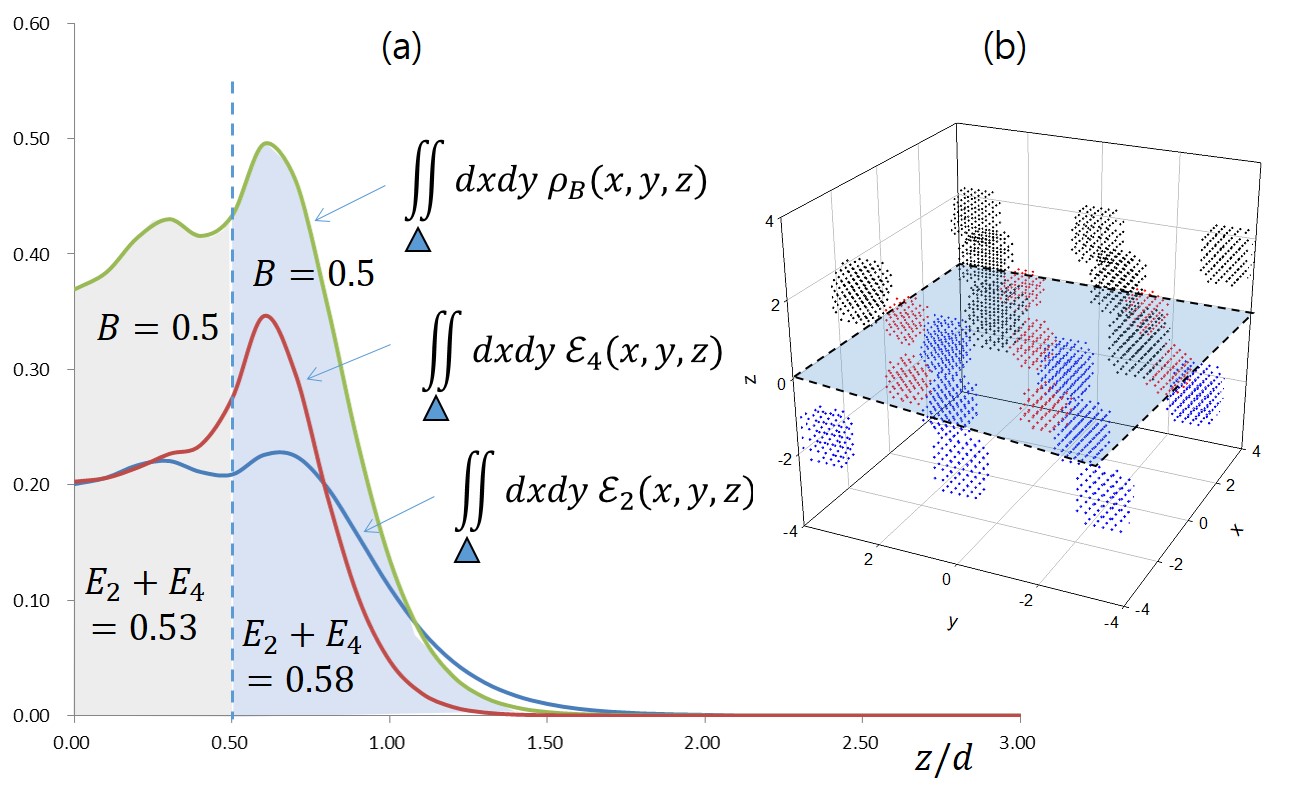}
  \caption{(a) Averaged densities over the triangular region as a function of $z$. (b) A rough three dimensional plot 
for the baryon number distribution of the minimum energy configuration for the ``doubly layered" skyrmion sheets. 
Actually, it is triple layers of ``half"-skyrmions.}
\end{center}
\end{figure}

\subsection{Inhomogeneous phase with a triple layer of skyrmion sheets}
The calculation follows a similar procedure as above.
Again, we fix the distance between the sheets to $d=L/\sqrt6$ and the lattice planes are $z=\pm d$ and $z=0$.   
The minimum energy configuration is obtained for ${\tilde\lambda}^2=0.053$ with $E/B=1.1$, which is slightly less than 
that of the double layer. The reason for this little reduction in $E/B$ is that the surface effect is still strong  in triple layer configuration.
That is, although a skyrmion in the middle layer has 12 nearest neighbors, the skyrmion in the outer layer has only 9. 

Shown in Fig.~8 are the average densities and a three dimensional plot for the configuration. 
Again, average densities outside of the lattice plane are almost the same as those of double layer and single sheet. 
This reminds us of the shape of heavy nuclei which have a similar density distribution near the surface 
and have the same surface thickness. 
The area below the curve tells us how much baryon number and energy are carried by matter in the hexagonal prism in a given range. 
Exactly half baryon number is lodged in each of the  three hexagonal prisms  $0\leq z\leq d/2$ and $d/2\leq z \leq d$ and $z\geq d$, 
while the energies carried by matter are 0.53, 0.53 and 0.58, in each location respectively.
This is exactly what we had expected. As shown naively in Fig.~8(b) half-skyrmions  have emerged. 
The half-skyrmions inside matter carry $E/B=1.06$ and those in the surface carry $E/B=1.16$. 
By generalizing this to any finite system, we can extract a mass formula which reads 
\begin{equation}
 E = 0.53 \times N_h^i + 0.56 \times N_h^s
\end{equation}
where $N_h^i$ is the number of half-skyrmions in the bulk of matter and $N_h^s$ denote those on the surface.

\begin{figure}
\begin{center}
  \includegraphics[width=0.7\linewidth]{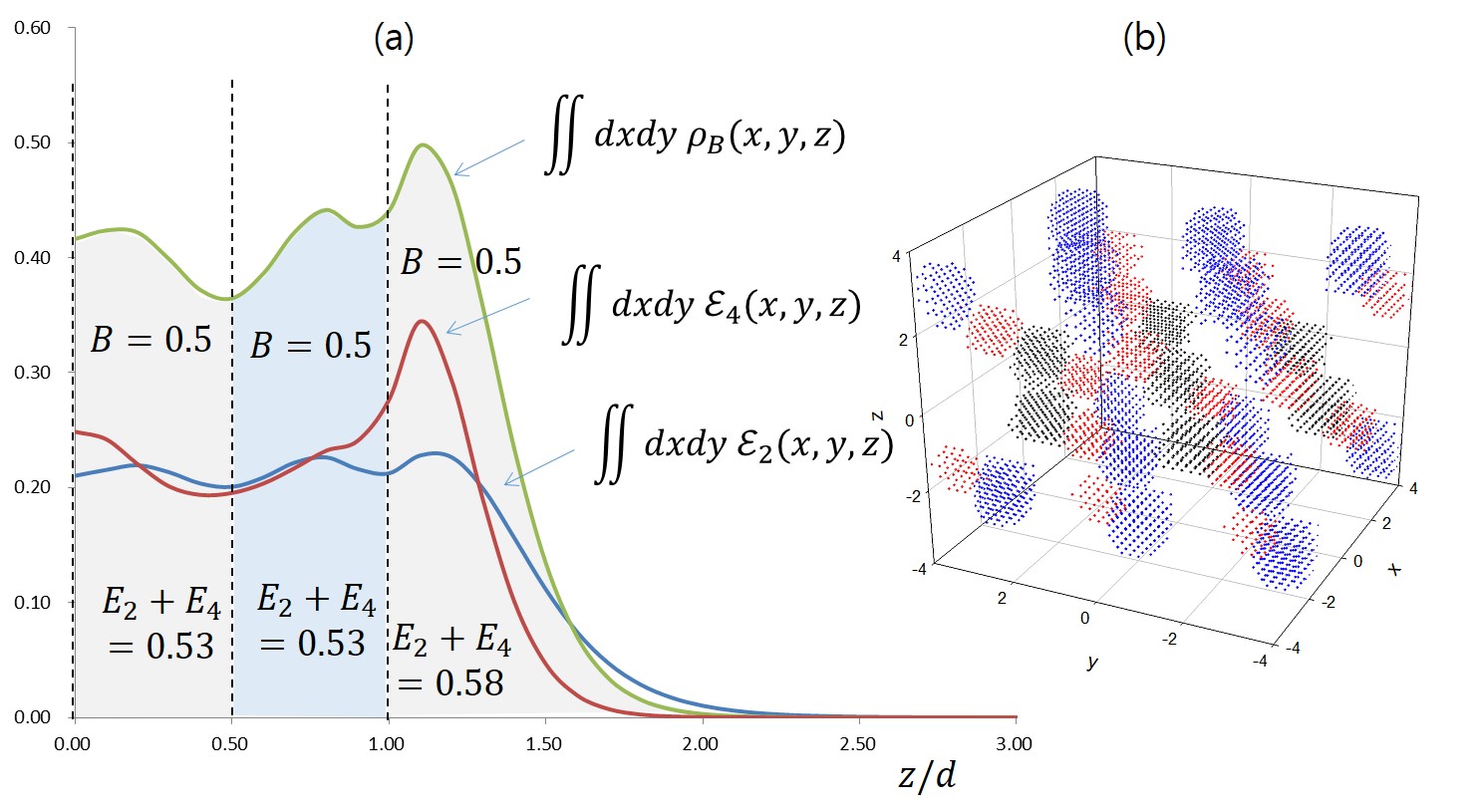}
  \caption{(a) averaged densities over the triangle as a function of $z$ and (b) a rough three diemsional plot for the triply layered skyrmion sheets.
Actual shape is five layers of half-skyrmion sheets. }
\end{center}
\end{figure}

\section{Conclusions}
It has been shown in the past that  the Skyrme model has a phase transition from an FCC crystal at low densities 
to a CC half-skyrmion crystal at high densities. At that time we pointed out that the FCC phase was unstable due to negative pressure and predicted
 that inhomogeneous phases in line of the Fermi drop model would be preferable. We have constructed such a phase in here which resembles very much
 some of the phases suggested for nuclear matter in the core of nuclear stars. In our case it is a phase made of matter sheets, i.e. a Lasagna type phase. We have been able to 
 construct this phase mathematically by means of the Atiyah-Manton ansatz using  the 't Hooft instanton function showing that its energy is lower than that of the FCC crystal and with a 
 characteristic  mass distribution whose support is a planar hexagonal lattice.
 
 This model for the low energy phase satisfies properties similar to those of conventional nuclear matter: it develops a skin, a skin depth and a bulk, with
 similar profiles as conventional nuclear matter. It leads to a mass formula which resembles a simplified version of the Weisz\"acker semi-empirical mass formula. Most important dynamically
 this result implies that our skyrmion system shares a common behavior to systems of particles interacting through strong  short-range forces.
 
A careful analysis of our calculation shows  that the long-range instanton profile function through its long range tails transforms into a short-range interaction among the skyrmions.
 This phenomenon is even more surprising since in our lagrangian the pion is massless. This observation seems to indicate that the short-range character of the strong nuclear force 
 is not crucially dependent on the mass of the pion but arises from confinement which manifests itself though the chiral SU(2) structure of the effective lagrangian. The Skyrme model is
 describing in an effective way how the massless gluons of QCD through confinement can give rise to manifestations of a strong short-range interaction
 with the limitations associated to the  large $N_c$ limit.

\section*{Acknowledgements}
VV acknoledges the hospitality and support of APCTP where these ideas were first discussed during the 12th APCTP-BLTP meeting. This work was supported in part by MICINN and UE Feder under contract FPA2016-77177-C2-1-P and SEV-2014-0398.

\end{document}